\documentclass[universe,communication,accept,moreauthors,pdftex,10pt,a4paper]{Definitions/mdpi}

\usepackage{dsfont}
\usepackage{url}

\firstpage{1} 
\pubvolume{xx}
\issuenum{1}
\articlenumber{5}
\pubyear{2019}
\copyrightyear{2019}
\history{Received: date; Accepted: date; Published: date}

\Title{Compact star properties from an extended linear sigma model}

\Author{János Takátsy $^{1,*}$, Péter Kovács $^{2}$, Zsolt Szép $^{3}$ and György Wolf $^{2}$}

\AuthorNames{Firstname Lastname, Firstname Lastname and Firstname Lastname}

\address{%
$^{1}$ Institute of Physics, ELTE E\"{o}tv\"{o}s University, H-1117 Budapest, Hungary\\
$^{2}$ Institute for Particle and Nuclear Physics, Wigner Research Centre for Physics, Hungarian Academy of Sciences, H-1525 Budapest, Hungary\\
$^{3}$ MTA-ELTE Theoretical Physics Research Group, H-1117 Budapest, Hungary}

\corres{takatsyj@caesar.elte.hu}

\firstnote{This paper is based on the talk at the 18th Zimányi School, Budapest, Hungary, 3–7 Dec. 2018.} 

\abstract{The equation of state provided by effective models of strongly interacting matter should comply with the restrictions imposed by current astrophysical observations of compact stars. Using the equation of state given by the (axial-)vector meson extended linear sigma model, we determine the mass-radius relation and study whether these restrictions are satisfied under the assumption that most of the star is filled with quark matter. We also compare the mass-radius sequence with those given by the equations of state of somewhat simpler models.}

\begin{document}
\section{Introduction}

A lot of theoretical and experimental effort is devoted to study the strong interaction under extreme conditions. The experiments ALICE \cite{ALICE} at CERN, and PHENIX \cite{PHENIX} and STAR \cite{Tlusty:2018rif} at RHIC explored the strongly interacting matter at low density and high temperature. In this region the situation is also satisfactory on the theoretical side, however, lattice calculations applicable at low density cannot yet be used at high densities \cite{Borsanyi:2016bzg}. Hence, effective models are needed in the high density region where the existing experimental data (NA61 \cite{NA61} at CERN, BES/STAR \cite{Tlusty:2018rif} at RHIC) is scarce and have rather bad statistics. Soon to be finished experimental facilities (NICA \cite{Kekelidze:2017ual} at JINR and CBM \cite{Ablyazimov:2017guv} at FAIR) are designed to explore this region more precisely.

For studying the cold, high density matter, new experimental information emerged in the past decade in a region of the phase diagram that is inaccessible to terrestrial experiments: the properties of neutron stars \cite{lattimer2007,Watts:2016uzu,Tews:2018chv}. Since the Tolman-Oppenheimer-Volkoff (TOV) equations \cite{Tolman, OppVolk} provide a direct relation between the equation of state (EoS) of the compact star matter and the mass-radius (M-R) relation of the compact star, these data can help to select those effective models, used to describe the strongly interacting matter, whose predictions are consistent with compact star observables. For example, the EoS must support the existence of a two-solar-mass neutron star \cite{Demorest,Antoniadis}. For the radius we have less stringent constraints. Bayesian analyses provide some window for the probable values for compact star radii \cite{Hell:2014xva,lattimer2014,Blaschke,Rezzolla,Abbott2018}. Based on these studies, in this paper we are adopting a radius window of $11.0$-$12.5$~km for compact stars with masses of $2$~$M_\odot$. The NICER experiment \cite{Watts:2016uzu,Watts2019} will provide very precise data on the masses and radii of neutron stars simultaneously.

Based on the above considerations, we investigate mass-radius sequences given by the EoS obtained in \cite{kovacs2016} from the $N_f=2+1$ flavor extended linear sigma model introduced in \cite{parganlija2013}. The model used here should be undoubtedly regarded in this context as a very crude approximation and the present work has to be considered only as our first attempt to study the problem. This is because the model, which is built on the chiral symmetry of the QCD, contains constituent quarks and therefore does not describe a realistic nuclear matter which is expected to form the crust of the compact star. The model is only applicable to some extent under the assumptions that at high densities nucleons dissolve into a see of quarks and a large part of the compact star is in that state. In other words we investigate here a quark star instead of a neutron or a hybrid star, which would be more realistic. The study in \cite{Zacchi:2015lwa} showed that a pure quark star of mass $\sim2$~$M_\odot$ can be achieved in a mean-field treatment of the $N_f=2+1$ linear sigma model if the Yukawa coupling between vector and quark fields is large enough. In the two flavor Nambu–Jona-Lasinio model, inclusion of 8-order quark interaction in the vector coupling channel also resulted in the stiffening of the quark equation of state \cite{Benic:2014iaa,Benic:2014jia}. Recently the existence of quark-matter cores inside compact stars was investigated also in \cite{Hell:2014xva} and \cite{Annala:2019eax}. It was found in \cite{Hell:2014xva} within a hybrid star model -- in which the quark core was described with a three-flavor Polyakov–Nambu–Jona-Lasinio model -- that the current astrophysical constraints can be fulfilled provided the vector interaction is strong enough. While in \cite{Annala:2019eax} it is claimed that the existence of quark cores in case of EoSs permitted by observational constraints is a common feature and should not be regarded as a peculiarity.

The paper is organized as follows. In Section~\ref{sec:model} we present the model and discuss how its solution, obtained in \cite{kovacs2016}, which reproduced quite well some thermodynamic quantities measured on the lattice, can be used in the presence of a vector meson introduced here to realize the short-range repulsive interaction between quarks in the simplest possible way. In Section~\ref{sec:res} we compare our results for the EoS and the M-R relation (star sequences), obtained in the eLSM for various values of the vector coupling $g_v$, to results obtained in the two-flavor Walecka model and the three-flavor non-interacting quark model. We draw the conclusions in Section~\ref{sec:concl} and discuss possible ways to improve the treatment of the model.

\section{Methods \label{sec:model}}

The model used in this paper is an $N_f = 2+1$ flavor (axial)vector meson extended linear sigma model (eLSM). The Lagrangian and the detailed description of this model, in which in addition to the full nonets of (pseudo)scalar mesons the nonets of (axial)vector mesons are also included, can be found in \cite{kovacs2016,parganlija2013}. The model contains three flavors of constituent quarks, with kinetic terms and Yukawa-type interactions with the (pseudo)scalar mesons. An explicit symmetry breaking of the mesonic potentials is realized by external fields, which results in two scalar expectation values, $\phi_\mathrm{N}$ and $\phi_\mathrm{S}.$\footnote{N and S denote the the non\,-\,strange and strange condensates, which are coupled to the 3x3 matrices $\lambda_\mathrm{N}=(\sqrt{2}\lambda_0+\lambda_8)/\sqrt{3}$ and $\lambda_\mathrm{S}=(\lambda_0-\sqrt{2}\lambda_8)/\sqrt{3}$, with $\lambda_8$ being the 8th Gell-Mann matrix and $\lambda_0=\sqrt{2/3}\mathds{1}$.} 

Compared to \cite{kovacs2016}, the only modification to the model is that we include in the Lagrangian a Yukawa term $-g_v\sqrt{6}\bar\Psi \gamma_\mu V_0^\mu\Psi$, which couples the quark field $\Psi^{\rm T}=(u,d,s)$ to the $U_V(1)$ symmetric vector field, that is $V_0^\mu = \frac{1}{\sqrt{6}} \mathrm{diag}(v_0 + \frac{v_8}{\sqrt{2}}, v_0 + \frac{v_8}{\sqrt{2}}, v_0-\sqrt{2}v_8)^\mu$. The vector meson field is treated at the mean-field level as in the Walecka model \cite{walecka}, but as a simplification we assign a nonzero expectation value only to $v_0^0$: $v_0^\mu\to v_0\delta^{0\mu}$ and $v_8^\mu\to 0$. Although this assignment is not physical, in this way the chemical potentials of all three quarks are shifted by the same amount, allowing us to use, as shown below, the result obtained in \cite{kovacs2016}. With the parameters used in \cite{kovacs2016}, the mass of the vector meson $v_0^\mu$ turns out to be $m_v=871.9$~MeV.

Since a compact star is relatively cold ($T\approx 0.1$~keV), we work at $T=0$~MeV using the approximation employed in \cite{kovacs2016}. We have three background fields $\phi_\mathrm{N},\phi_\mathrm{S}$ and $v_0$, and the calculation of the grand potential, $\Omega$, is performed using a mean-field approximation, in which fermionic fluctuations are included at one-loop order, while the mesons are treated at tree-level. Hence, the grand potential can be written in the following form:
\begin{equation}
\label{Eq:grand_pot}
    \Omega(\mu_q; \phi_\mathrm{N}, \phi_\mathrm{S},v_0) = U_\mathrm{mes}(\phi_\mathrm{N},\phi_\mathrm{S}) - \frac{1}{2}m_v^2 v_0^2 + \Omega^{(0)\mathrm{vac}}_{q\bar{q}}(\phi_\mathrm{N}, \phi_\mathrm{S}) 
    + \Omega^{(0)\mathrm{matter}}_{q\bar{q}}(\tilde\mu_q;\phi_\mathrm{N}, \phi_\mathrm{S}) \:,
\end{equation}
where $\tilde\mu_q=\mu_q-g_v v_0$ is the effective chemical potential of the quarks, while $\mu_q = \mu_\mathrm{B}/3$ is the physical quark chemical potential, with $\mu_B$ being the baryochemical potential. On the right hand side of the grand potential \eqref{Eq:grand_pot}, the terms are (from left to right): the tree-level potential of the scalar mesons, the tree-level contribution of the vector meson, the vacuum and the matter part of the fermionic contribution at vanishing mesonic fluctuating fields. The fermionic part is obtained by integrating out the quark fields in the partition function. The vacuum part was renormalized at the scale $M_0=351$~MeV. More details on the derivation can be found in \cite{kovacs2016}.

The background fields $\phi_\mathrm{N}, \phi_\mathrm{S}$, and $v_0$ are determined from the stationary
conditions
\begin{equation}
    \frac{\partial \Omega}{\partial \phi_\mathrm{N}}\bigg|_{\phi_\mathrm{N}=\bar\phi_\mathrm{N}} = \frac{\partial \Omega}{\partial \phi_\mathrm{S}}\bigg|_{\phi_\mathrm{S}=\bar\phi_\mathrm{S}}  = 0 \quad\textnormal{and} \quad \frac{\partial \Omega}{\partial v_0}\bigg|_{v_0=\bar{v}_0} =0 \: ,
    \label{Eq:stat_cond}
\end{equation}
where the solution is indicated with a bar. Since $\partial/\partial v_0=-g_v \partial/\partial{\tilde \mu_q}$, the stationary condition with respect to $v_0$ reads 
\begin{equation}
\label{Eq:gap-w0}
\bar{v}_0(\phi_\mathrm{N},\phi_\mathrm{S}) = \frac{g_v}{m_v^2} \rho_q(\tilde\mu_q(\bar{v}_0);\phi_\mathrm{N},\phi_\mathrm{S}) \:,
\end{equation}
where $\rho_q(x;\phi_\mathrm{N},\phi_\mathrm{S})=-\partial\Omega^{(0)\mathrm{matter}}_{q\bar{q}}(x;\phi_\mathrm{N},\phi_\mathrm{S})/\partial x.$

 When solving the model, we give values to $g_v$ in the range $[0,3)$, while for the remaining 14 parameters of the model Lagrangian we use the values given in Table IV of \cite{kovacs2016}. These values were determined there by calculating constituent quark masses, (pseudo)scalar curvature masses with fermionic contribution included and decay widths at $T=\mu_q=0$ and comparing them to their experimental PDG values \cite{PDG}. Parameter fitting was done using a multiparametric $\chi^2$ minimization procedure \cite{james1975}. In addition to the vacuum quantities, the pseudocritical temperature $T_\mathrm{pc}$ at $\mu_q=0$ was also fitted to the corresponding lattice result \cite{aoki2006, Bazavov:2011nk}. We mention that the model also contains the Polyakov-loop degrees of freedom (see \cite{kovacs2016} for details), but to keep the presentation simple we omitted them from \eqref{Eq:grand_pot}, as at $T=0$ they do not contribute to the EoS directly. Their influence is only through the value of the model parameters taken from \cite{kovacs2016}: since they modify the Fermi-Dirac distribution function, they influence the value of $T_\mathrm{pc}$ used for parameterization, as described above. A parameterization based on vacuum quantities alone could lead to unphysically large values of $T_\mathrm{pc}$ and, compared to the case when $T_\mathrm{pc}$ is included in the fit, also to different assignments of scalar nonet states to physical particles, that is $\chi^2$ could become minimal for a different particle assignment.
 
 The solution of the model at $g_v=0$, obtained in \cite{kovacs2016}, can be used to construct the solution at $g_v\ne0$ (see {\it e.g.} Ch. 2.1 of \cite{almasi}): one only has to interpret the solution at $g_v=0$ as a solution obtained at a given $\tilde \mu_q$ and determine $\mu_q$ at some $g_v\ne 0$ using \eqref{Eq:gap-w0}. To see that the solutions $\bar\phi_\mathrm{N,S}$ for $g_v\ne0$ can be related to the solution obtained at $g_v=0,$ where $\bar{v}_0=0$, consider the grand potential at $g_v=0$. This potential, denoted as $\Omega_0$, is subject to the stationary conditions \eqref{Eq:stat_cond} with solutions $\bar\phi_\mathrm{N,S}^0(\mu_q)$. It is then easy to see using \eqref{Eq:grand_pot}, that the solution $\bar\phi_\mathrm{N,S}(\mu_q)$ of \eqref{Eq:stat_cond} satisfies $\bar\phi_\mathrm{N,S}(\mu_q+g_v v_0)=\bar\phi_\mathrm{N,S}^0(\mu_q)$ or, changing the variable $\mu_q$ to $\tilde \mu_q$ the relation becomes 
\begin{equation}
\bar\phi_\mathrm{N,S}(\tilde\mu_q+g_v v_0)=\bar\phi_\mathrm{N,S}^0(\tilde\mu_q).
\end{equation}
The value of the grand potential $\Omega$ at the extremum can be given in terms of the value of the grand potential with $g_v=0$, that is $\Omega_0$, at its extremum. Using that the extrema of $\Omega_0(\tilde\mu_q,\phi_\mathrm{N},\phi_\mathrm{S},v_0=0)$ are $\bar\phi_\mathrm{N}^0$ and $\bar\phi_\mathrm{S}^0$, one has
\begin{equation}
\Omega(\mu_q;\bar\phi_\mathrm{N}(\mu_q), \bar\phi_\mathrm{S}(\mu_q), \bar{v}_0) = \Omega_0(\tilde\mu_q,\phi_\mathrm{N}^0(\tilde\mu_q),\phi_\mathrm{S}^0(\tilde\mu_q),v_0=0)-\frac{1}{2}m^2_v\bar{v}_0^2\: ,
\end{equation}
where $\bar{v}_0\equiv\bar{v}_0(\bar\phi_\mathrm{N}(\mu_q),\bar\phi_\mathrm{S}(\mu_q))=\frac{g_v}{m^2_v} \rho_q(\tilde\mu_q; \bar\phi_\mathrm{N}(\mu_q), \bar\phi_\mathrm{S}(\mu_q))=\frac{g_v}{m^2_v} \rho_q(\tilde\mu_q; \bar\phi_\mathrm{N}^0(\tilde\mu_q), \bar\phi_\mathrm{S}^0(\tilde\mu_q))$ and $\mu_q=\tilde\mu_q+g_v \bar{v}_0$.

The pressure $p$ and the energy density $\varepsilon$ are calculated from the grand potential. At $v_0\ne0$ they can be expressed in terms on the pressure obtained at $g_v=0$: 
\begin{eqnarray}
    p(\mu_q) &=& \Omega(\mu_q=0;\bar\phi_\mathrm{N}(0),\bar\phi_\mathrm{S}(0),\bar{v}_0(0)) 
    - \Omega(\mu_q; \bar\phi_\mathrm{N}, \bar\phi_\mathrm{S},\bar{v}_0)\nonumber\\
    &=& \Omega_0(\tilde\mu_q=0;\bar\phi_\mathrm{N}^0(0),\bar\phi_\mathrm{S}^0(0),v_0=0) 
    - \Omega_0(\tilde\mu_q; \bar\phi_\mathrm{N}^0, \bar\phi_\mathrm{S}^0,v_0=0)
    +\frac{1}{2} m^2_v\bar{v}_0^2\nonumber\\ 
    &=& p(\tilde\mu_q)|_{g_v=0} + \frac{1}{2} m^2_v\bar{v}_0^2\: ,
\end{eqnarray}
where $\bar{v}_0=\frac{g_v}{m^2_v} \rho_q(\tilde\mu_q; \bar\phi_\mathrm{N}^0(\tilde\mu_q), \bar\phi_\mathrm{S}^0(\tilde\mu_q))$, and then $\varepsilon = -p + \mu_q \rho_q$, where $\mu_q=\tilde\mu_q+g_v\bar{v}_0$.

With the EoS $p(\varepsilon)$ obtained at $T=0$ and high densities we determine the mass-radius relation of non-rotating static compact stars by solving the TOV equation \cite{Tolman,OppVolk} using a fourth-order Runge-Kutta differential equation integrator with adaptive stepsize-control.

\section{Results \label{sec:res}}

Since our eLSM model was fitted to the hadron spectrum and not to the nuclear matter, we compare its results with those obtained in two relativistic models generally used in the description of compact stars, in order to assess the importance of various ingredients involved in these models. For comparison we consider the three-flavor non-interacting constituent quark model (see {\it e.g.} \cite{schmitt} and \cite{glendenning}) and the Walecka model, which in its simplest form contains the proton and neutron, the scalar-isoscalar meson $\sigma$ and the isoscalar-vector meson $\omega$ \cite{walecka}. The use of the Walecka model for the description of the neutron stars requires charge neutrality, which calls for the introduction of the $\rho$ meson in order to have a proper description of the nuclear symmetry energy \cite{glendenning}.

In the non-interacting constituent quark model the masses are fixed to \mbox{$m_u=m_d=75$ MeV,} $m_s=365$ MeV, values obtained from our eLSM at the 1st order chiral phase transition point, that is at $\mu_{q,c}\approx 323$~MeV, where the potential is degenerate. The calculation of the energy density and pressure was done with the bag constant $B^{1/4}=163$~MeV. Including electrons in the model, the conditions of $\beta$-equilibrium and charge neutrality were taken into account. 

We use two mean-field versions of the Walecka model, one that includes the effect of the scalar self-interaction through a classical potential with cubic and quartic terms of the form
\begin{equation}
V_{I,\sigma}=\frac{b}{3} m_N(g_\sigma \sigma)^3 + \frac{c}{4}(g_\sigma \sigma)^4,
\end{equation}
and a version where the scalar self-interaction is neglected. Using $m_\sigma=550$~MeV, $m_\omega=783$~MeV and $m_\rho=775.3$~MeV for the mesons and $m_N=939$~MeV for the nucleon mass, the parameters are fixed from nuclear matter properties: the value $n_0=0.153~\textnormal{fm}^{-3}$ for the saturation density (where $p=0$), the nuclear binding energy per nucleon $E_0=(\varepsilon/n_0-m_N)=-16.3~$~MeV and the symmetry energy coefficient, for which we take the value $a_\textnormal{sym}=31.3$~MeV \cite{Xu:2010fh}, and in the version with scalar self-interactions also the compression modulus $K=250$~MeV and the Landau mass $m_L=0.83m_N$. The values of the parameters used here are basically those of \cite{schmitt}: for the value of the Yukawa couplings of the mesons to the nucleons one has $g^2_\sigma=9.5372/(4\pi)$, $g^2_\omega=14.717/(4\pi)$ and $g_\rho=6.8872$ when the scalar self-interaction is neglected, while in the other case $g^2_\sigma=6.003/(4\pi)$, $g^2_\omega=5.9484/(4\pi)$, $g_\rho=8.3235$, $b=7.95\cdot10^{-3}$ and $c=6.947 \cdot 10^{-4}$. For a recent study of the effect of $K$, $m_L$ and of the form of the scalar potential on the mass-radius relation we refer the interested reader to \cite{Posfay:2019nto}.


\begin{figure}[!t]
\centering
\includegraphics[height=0.39\linewidth]{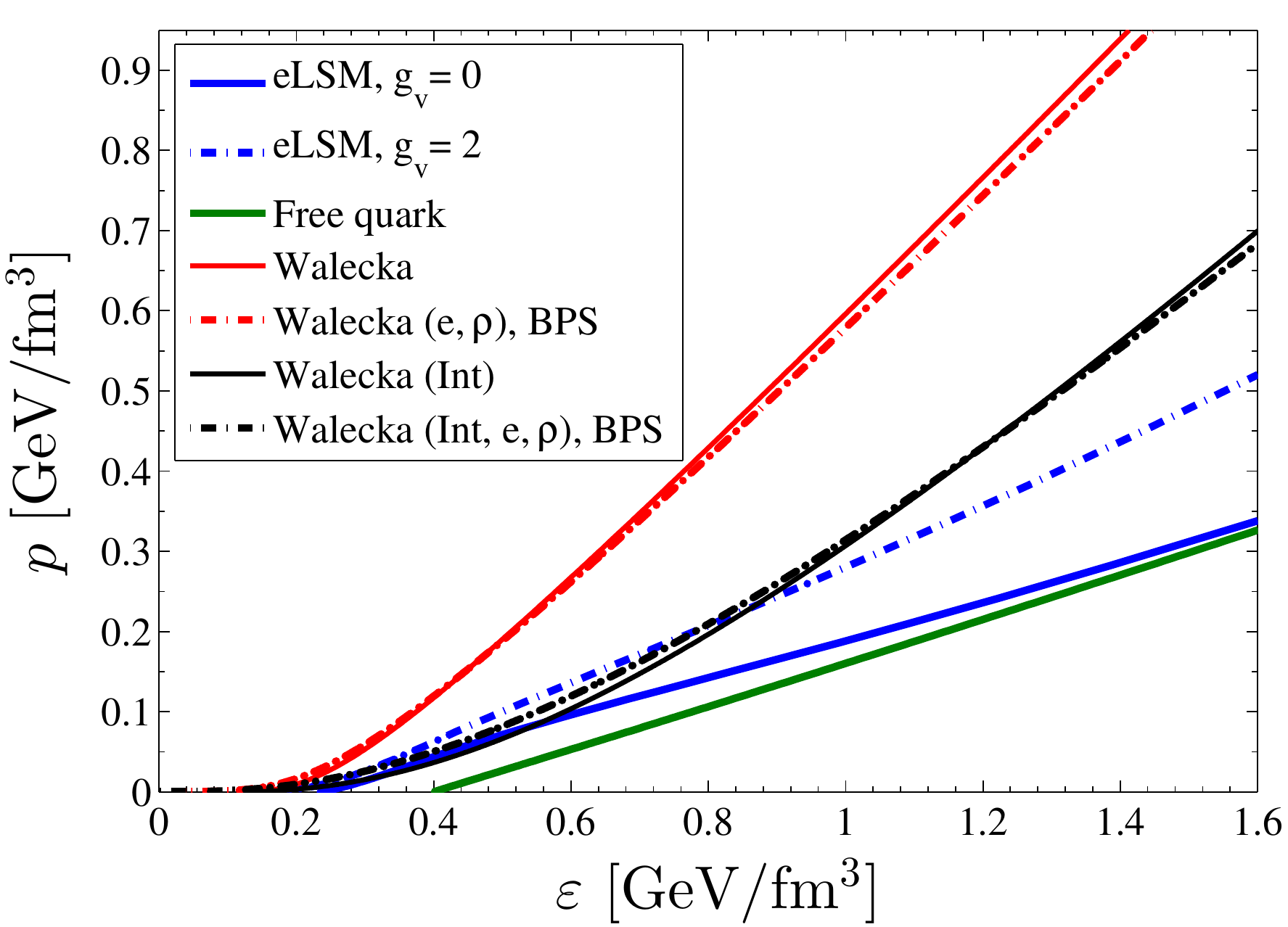}
\includegraphics[height=0.39\linewidth]{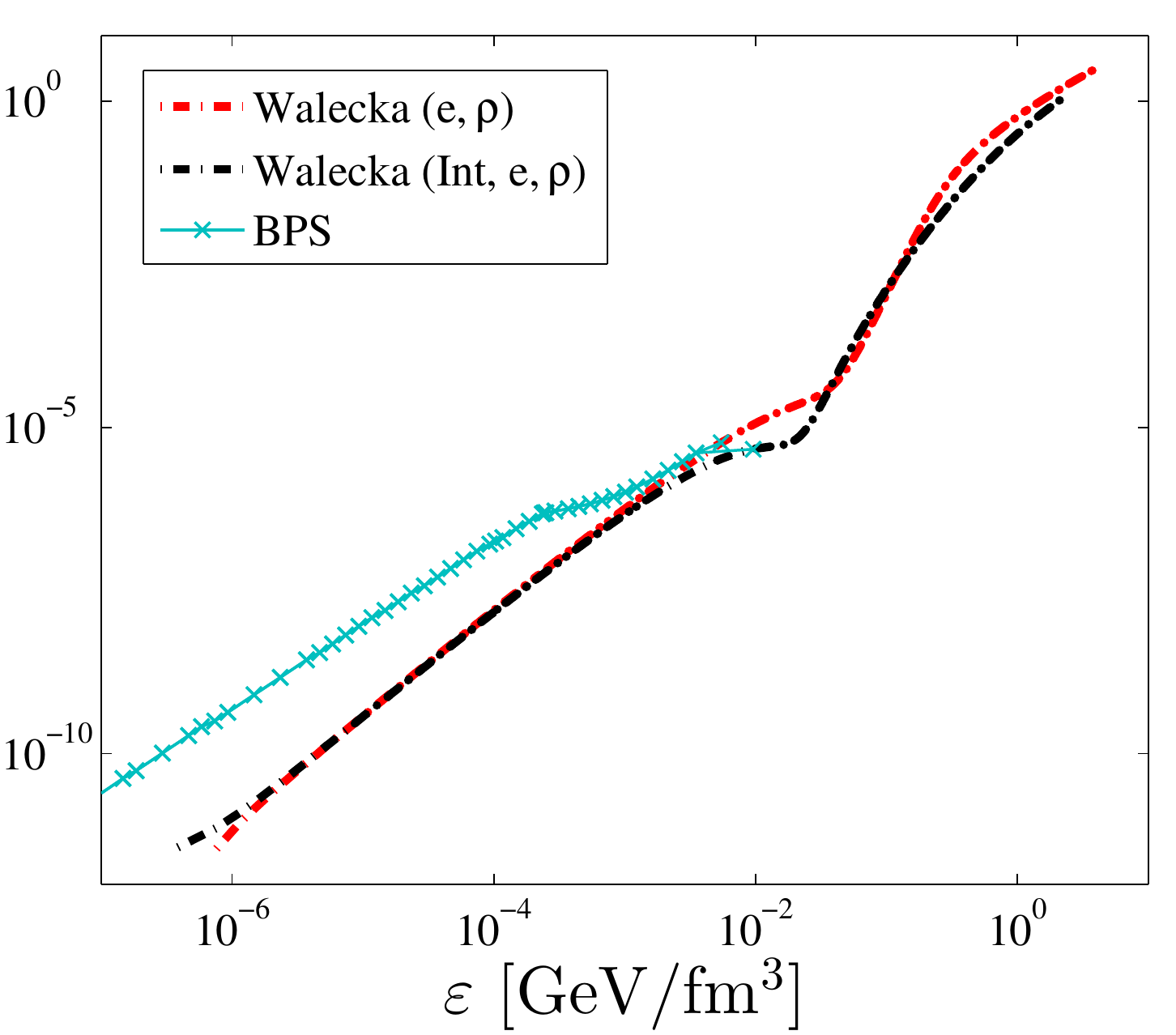}
\caption{Left panel: The $T=0$ EoS of the eLSM (blue solid line for $g_v=0$ and blue dashed-dotted line for $g_v=2$) compared to those of the free constituent quark matter with mass values given in the text (green solid line) and of the Walecka model with (black lines) and without (red lines) including the scalar self-inteaction. For the latter model the dashed-dotted line type indicates that $\beta$-equilibrium and charge neutrality are imposed, the $\rho$ meson is included and that at low energy densities the EoS is replaced by the BPS EoS. Right panel: matching the EoS of the Walecka model to the BPS EoS (see the text for details). Notice that the consequence of imposing the mentioned compact star constraints (inclusion of electrons and $\rho$) in the Walecka model is that $p(\varepsilon)>0$ even at low energy densities.}
\label{fig:EqoS}
\end{figure}

\begin{figure}[!b]
\centering
\includegraphics[width=0.49\linewidth]{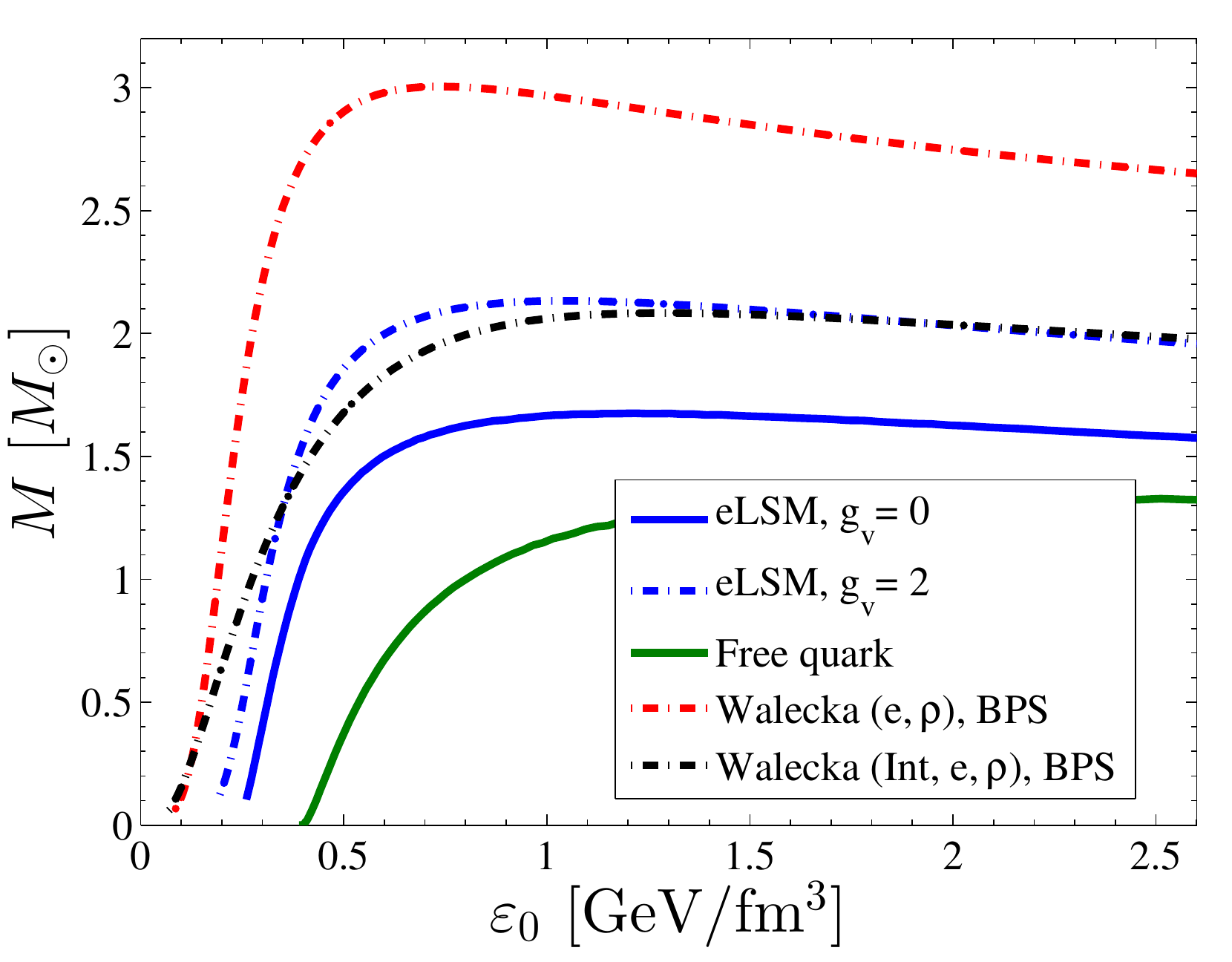}
\includegraphics[width=0.49\linewidth]{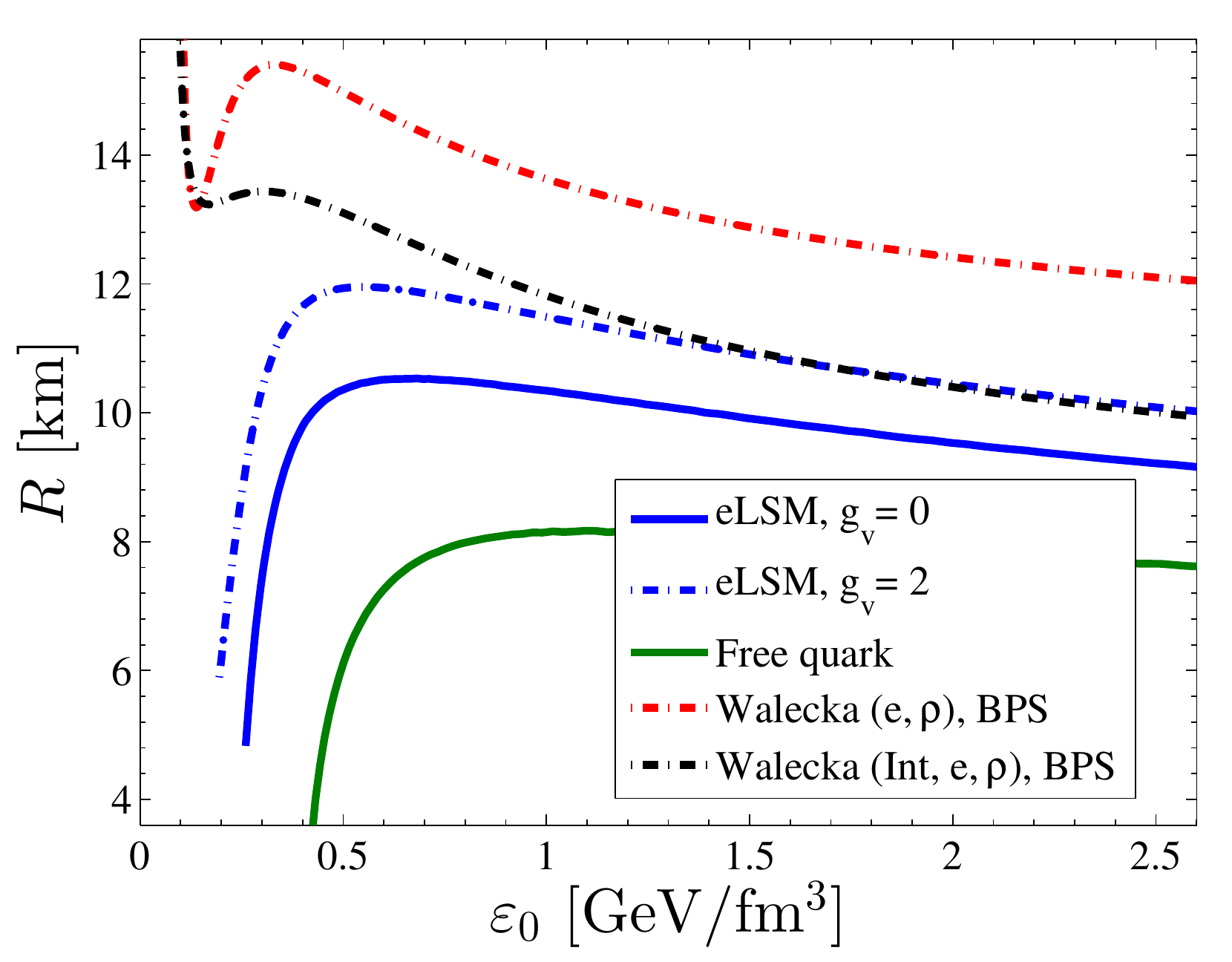}
\caption{The masses (left panel) and radii (right panel) of the compact stars as functions of the central energy density ($\varepsilon_0$). The line style for the different cases correspond to that of Fig.~\ref{fig:EqoS}.
}
\label{fig:epsRM}
\end{figure}

The EoS of the Walecka model subject to the constraints of $\beta$-equilibrium and charge neutrality is applicable only to the core of the compact star. A proper phenomenological description requires the modeling of the stellar matter in the crust and of the crust-core transition. In the present work we only implement, using the tabulated data from Tables 5.7 of \cite{glendenning}, the BPS EoS \cite{Baym:1971pw} for the outer crust of a neutron star whose core is described by the EoS of the Walecka model. This is done by simply replacing at low energy densities, corresponding to densities below the neutron drip line, $\rho_B\le0.01\mathrm{fm}^{-3}$, the EoS of the Walecka model with the BPS EoS, as indicated in the right panel of Fig.~\ref{fig:EqoS}. More sophisticated procedures for core-crust matching are described in \cite{Fortin:2016hny} together with their influence on the $M(R)$ relation. A realistic description of astrophysical data would require an additional matching to an EoS for the inner crust that applies for densities above the neutron drip density. This is beyond the scope of our present study and we refer the interested reader to a recent review \cite{Blaschke:2018mqw} that provides a detailed discussion of the neutron star crust matter and of the EoS of dense neutron star matter. Convenient analytic parameterizations of unified EoSs derived from a single model and describing the crust and the core of the neutron star are given in \cite{Potekhin:2013qqa}.

\begin{figure}[!t]
\centering
\includegraphics[width=0.7\linewidth]{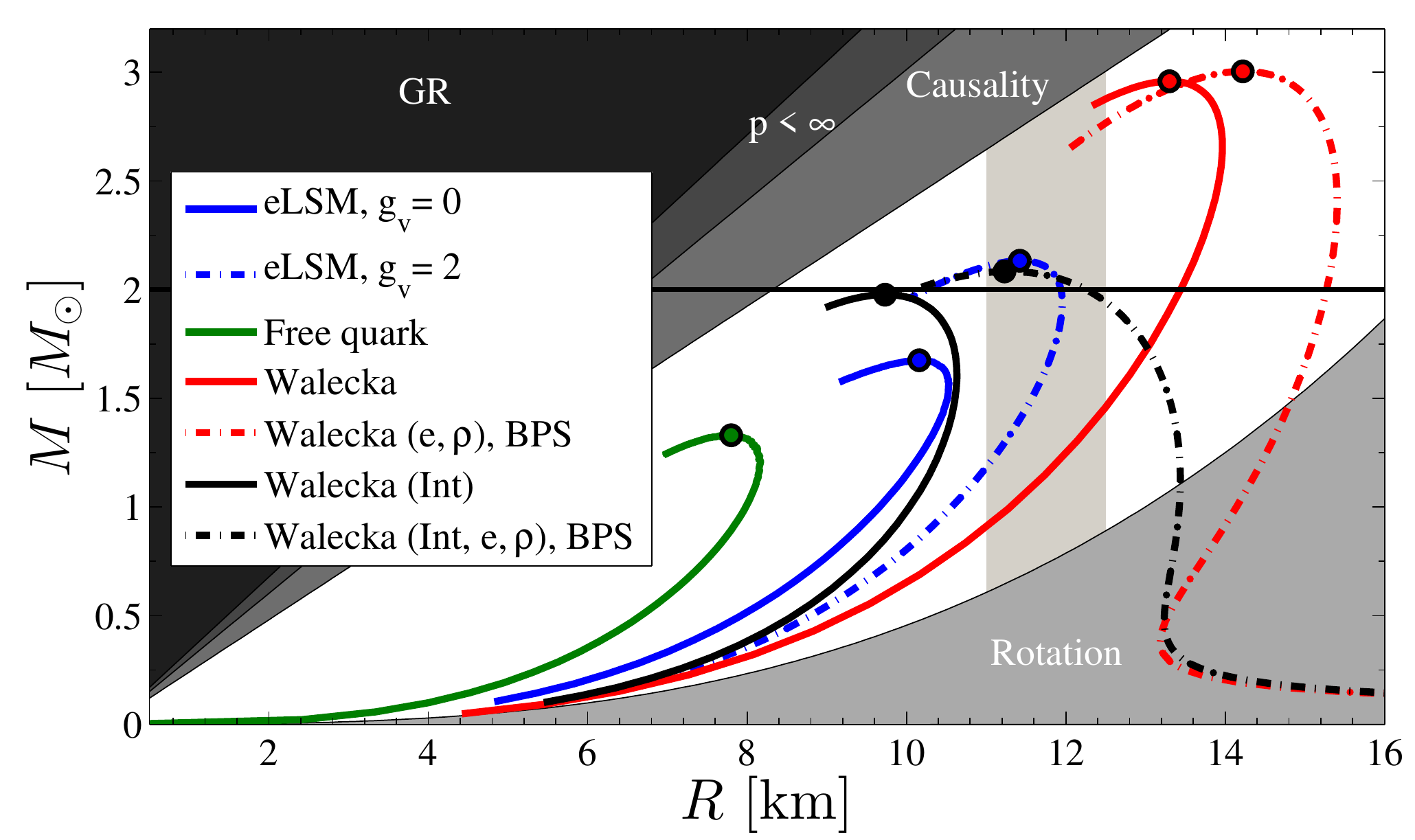}
\caption{Mass-radius relations for the eLSM (blue solid line for $g_v=0$ and blue dashed-dotted line for $g_v=2$), the free constituent quark matter (green solid line), and the Walecka model for the various cases of Fig.~\ref{fig:EqoS}. The ends of the stable sequences of compact stars are marked by blobs. The observational constraint set by observed pulsars with masses of $\sim2$~$M_\odot$ is represented by the black horizontal line, and the applied radius window of $11.0$-$12.5$~km at $2$~$M_\odot$ is depicted by the vertical shaded area. The different shaded regions are excluded by the GR constraint $R>2GM/c^2$, the finite pressure constraint $R>(9/4)GM/c^2$, causality $R>2.9GM/c^2$ and the rotational constraint based on the $716$~Hz pulsar J1748-2446ad, $M/M_\odot>4.6\cdot10^{-4} \: (R/\mathrm{km})^3$ \cite{hessels2006}. For a more detailed discussion on these constraints see {\it e.g.} \cite{lattimer2007}.}
\label{fig:MR}
\end{figure}

The zero-temperature EoSs are shown in Fig.~\ref{fig:EqoS}. For the Walecka model we also consider the case when charge neutrality condition and $\beta$-equilibrium with electrons are not imposed and the BPS EoS is not used. We can see in Fig.~\ref{fig:EqoS} that at small energy densities the pressure in the eLSM with $g_v=0$ is slightly higher than in the non-interacting quark model ({\it i.e.} the EoS is stiffer), but close to the value of the pressure obtained in the Walecka model with scalar self-interaction. This shows that the inclusion of scalar interactions in the Walecka model brings the EoS closer to that of the eLSM, as in case of the Walecka model the higher pressure corresponds to the non-interacting model. At high energy densities the values of the pressure in the eLSM with $g_v=0$ approach those obtained in the non-interacting quark model. Inclusion of the repulsive interaction between quarks in the eLSM renders the EoS stiffer compared to the $g_v=0$ case, as expected, and it brings the EoS of the eLSM closer to that obtained in the Walecka model with scalar self-interaction. It is worth noting that relatively small differences in the $p(\epsilon)$ lead to significant differences in the M-R curves, as we shall see later in Figs.~\ref{fig:epsRM} and \ref{fig:MR}. 

By solving the TOV equation using a specific EoS, one can obtain the radial dependence of the energy density (and thus of the pressure) for a certain central energy density, $\varepsilon_0$. One can then determine the mass and radius of the compact star for that central energy density. By changing $\varepsilon_0$, one gets a sequence of compact star masses and radii parameterized by the central energy density, as shown in Fig.~\ref{fig:epsRM} for various models. The sequence of stable compact stars ends when the maximum compact star mass is reached with increasing central energy density. 

\begin{figure}[!t]
\centering
\includegraphics[width=0.49\linewidth]{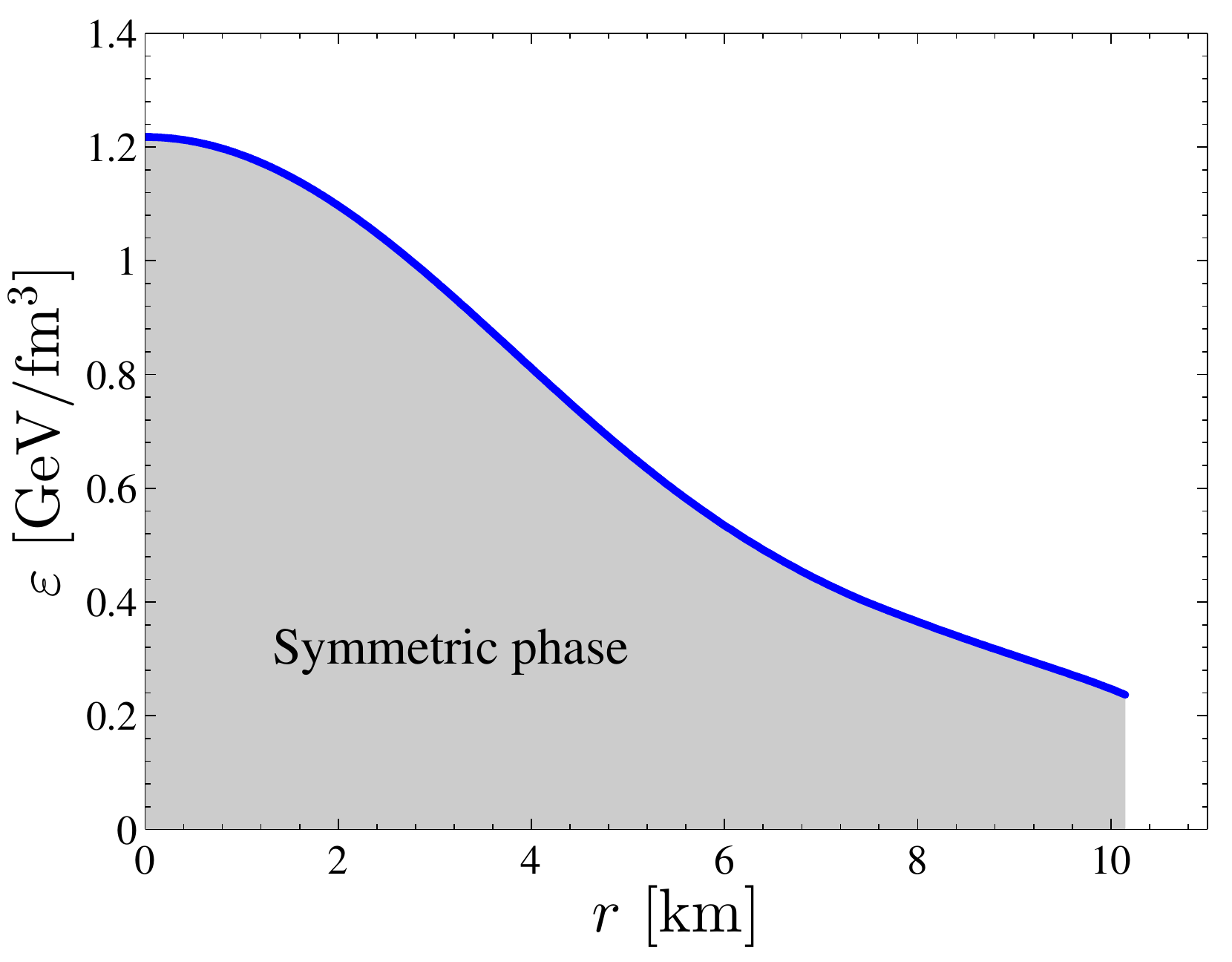}
\includegraphics[width=0.49\linewidth]{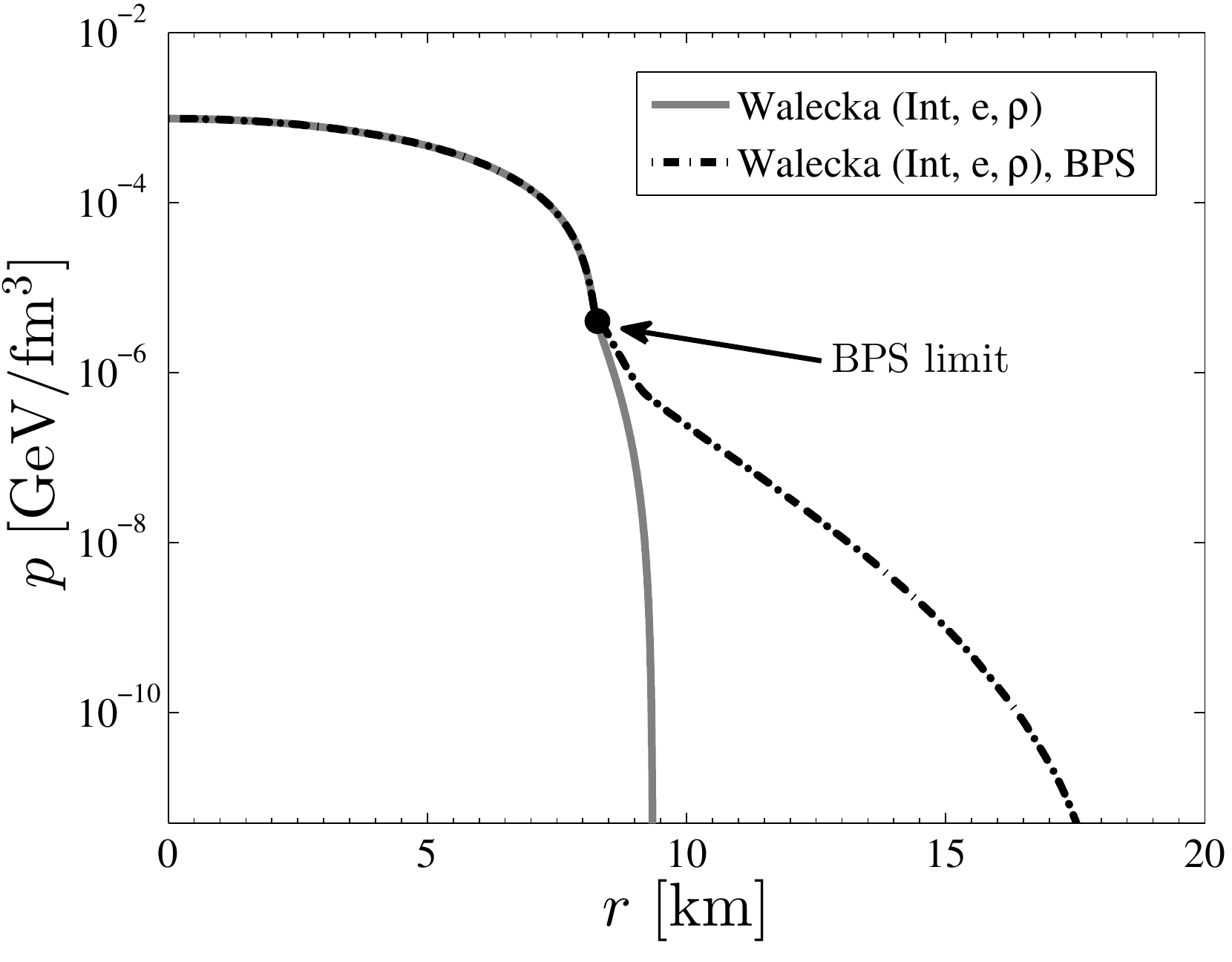}
\caption{Left panel: The energy density as a function of radial coordinate inside the maximum mass compact star corresponding to the eLSM with $g_v=0$. The chiral phase transition occurs at the very edge of the star, hence the whole star is basically composed of chirally symmetric quark matter. Right panel: To ilustrate the effect of the BPS EoS in the Walecka model, we show the pressure as a function of the radial distance for a central energy density of $\varepsilon_0=7\cdot 10^8$ MeV$^4$.}
\label{fig:epsR_maxmass}
\end{figure}

The mass-radius relations for the four models are shown in Fig.~\ref{fig:MR} together with physical constraints obtained from observations of binary pulsar systems and X-ray binaries. As expected based on Fig.~5.23 of \cite{glendenning}, the proper treatment of the neutron star outer crust by the BPS EoS that corresponds to a Coulomb lattice of different nuclei embedded in a gas of electrons has a remarkable influence on both the mass and the radius of the star (see also Fig.~\ref{fig:epsR_maxmass}): without the BPS EoS the turning point at the smallest radius of that part of the mass-radius diagram which corresponds to large stars with small masses is around $8$km ($9$km) in the Walecka model with (without) scalar self-interactions and the minimum mass of the stars with large radii is much smaller. Also, without the constraints of charge neutrality and $\beta$-equilibrium with electrons and without the effect of the $\rho$ meson, even the shape of the $M(R)$ curve obtained in the Walecka model is different.

As it can be seen in Fig.~\ref{fig:MR}, the maximum possible compact star mass is lower for models with less stiff EoSs. The star sequences corresponding to the non-interacting quark model, the Walecka model with scalar interactions, but without compact star constraints (no electrons and $\rho$ included) and the eLSM model without repulsive interaction in the vector sector does not lie in the desired radius window. The highest mass compact star has a mass of $\sim1.7$~$M_\odot$ for the eLSM with $g_v=0$, $\sim1.3$~$M_\odot$ for the non-interacting quark model, $\sim3$~$M_\odot$ for the non-interacting Walecka model, and $\sim2$~$M_\odot$ for the interacting Walecka model. It is interesting to note that the star sequence in the eLSM model without repulsive interaction is close to the one of the Walecka model with scalar interaction but without compact star constraints, although the later contains repulsive interaction as well. As expected, the repulsive interaction makes the EoS stiffer in the eLSM, and for $g_v=2$ a mass value of $\sim2.15$~$M_\odot$ can be reached with a radius at $M=2$~$M_\odot$ in the permitted radius window. Based on Fig.~\ref{fig:EqoS} one can observe that, interestingly, it is the stiffer EoS for $\varepsilon<0.8$ GeV/fm$^3$, as compared to the Walecka model including scalar interactions and not subject to compact star constraints, that brings the star sequence to the desired range in the eLSM with repulsive interaction.

\begin{figure}[!b]
\centering
\includegraphics[width=0.7\linewidth]{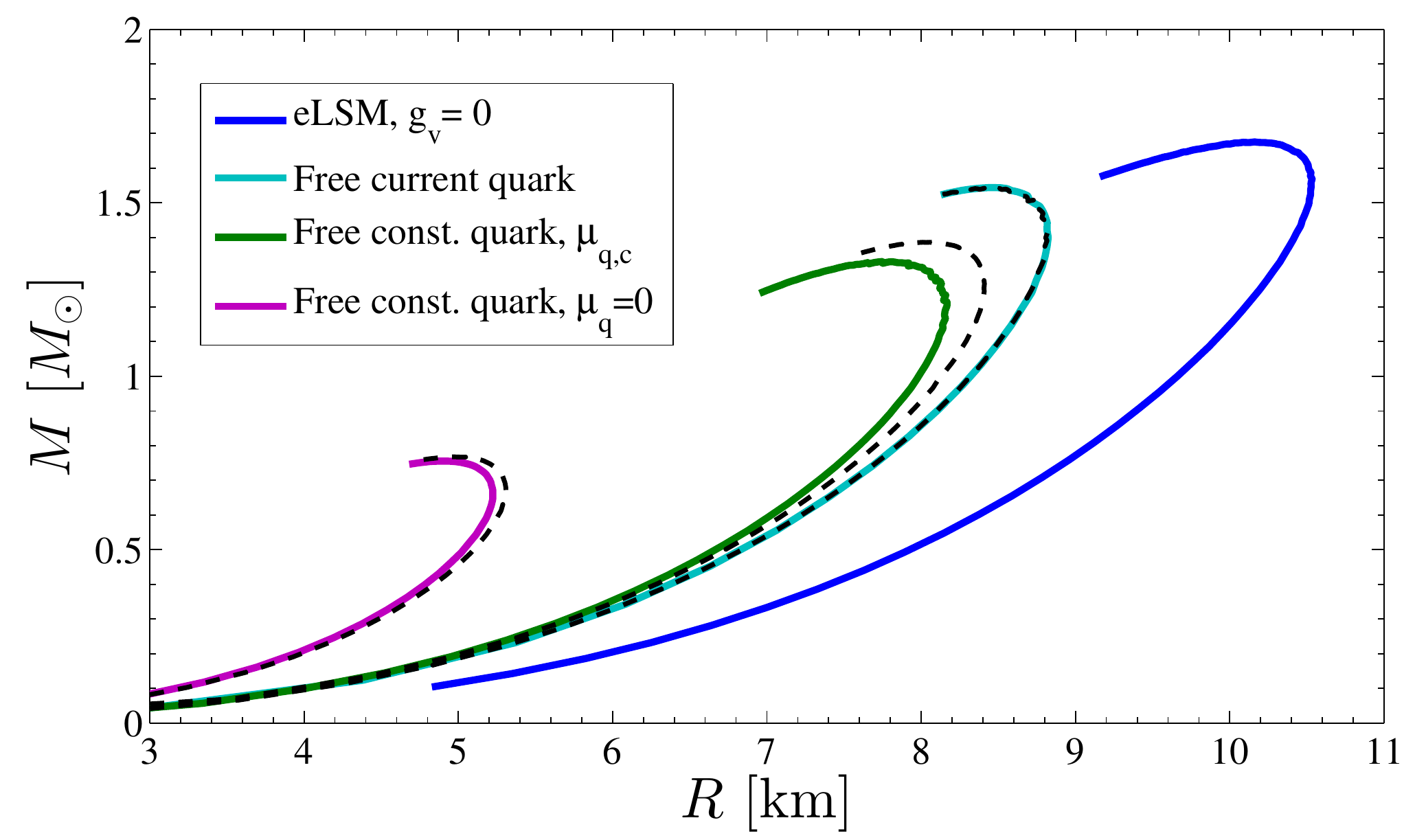}
\caption{$M(R)$ curves of the non-interacting quark model with three different quark mass setups (left curve: $m_u=m_d=322$~MeV, $m_s=458$~MeV, middle curve: $m_u=m_d=75$~MeV, $m_s=365$~MeV and right curve: $m_u=m_d=0$~MeV, $m_s=90$~MeV) compared to the $M(R)$ curve of the eLSM model with $g_v=0$ in which the quark masses change (rightmost curve). The dashed curves are obtained without imposing the constraints of charge neutrality and $\beta$-equilibrium.}
\label{fig:MR_quark}
\end{figure}

The energy density as a function of radial position is shown in Fig.~\ref{fig:epsR_maxmass} for the maximum mass compact star obtained with the EoS of the eLSM with $g_v=0$. Since the chiral phase transition occurs at $\mu_{q,c} \approx 323$~MeV, which essentially corresponds to zero pressure, almost all of the matter in the compact star is in the chirally symmetric phase ({\it i.e.} $\epsilon$ corresponds to $\mu_q>\mu_{q,c} \approx 323$~MeV). In the right panel we illustrate in the case of the Walecka model how the BPS EoS, which models the outer crust, influences the solution of the TOV equation.

In Fig.~\ref{fig:MR_quark} we compare $M(R)$ curves obtained in the non-interacting quark model at the three different sets of quark masses listed in the caption (two of them comes from the eLSM at the value of $\mu$ indicated in the key) and in the interacting eLSM model with $g_v=0$. For the free quark model the quark masses increase from right to left, as indicated in the caption, while in case of the eLSM the quark masses change (decreasing with increasing baryochemical potential) and their masses are smaller or equal than that of the leftmost curve and always larger than that of the rightmost curve obtained in the free quark model. This clearly shows the significant effect of interactions on the M-R curve. The dashed lines show that neglecting the constraints of charge neutrality and $\beta$-equilibrium in the non-interacting quark model does not lead to significant changes. Consequently, we also expect these constraints to have a mild effect in the eLSM.

\begin{figure}[!t]
\centering
\includegraphics[width=0.7\linewidth]{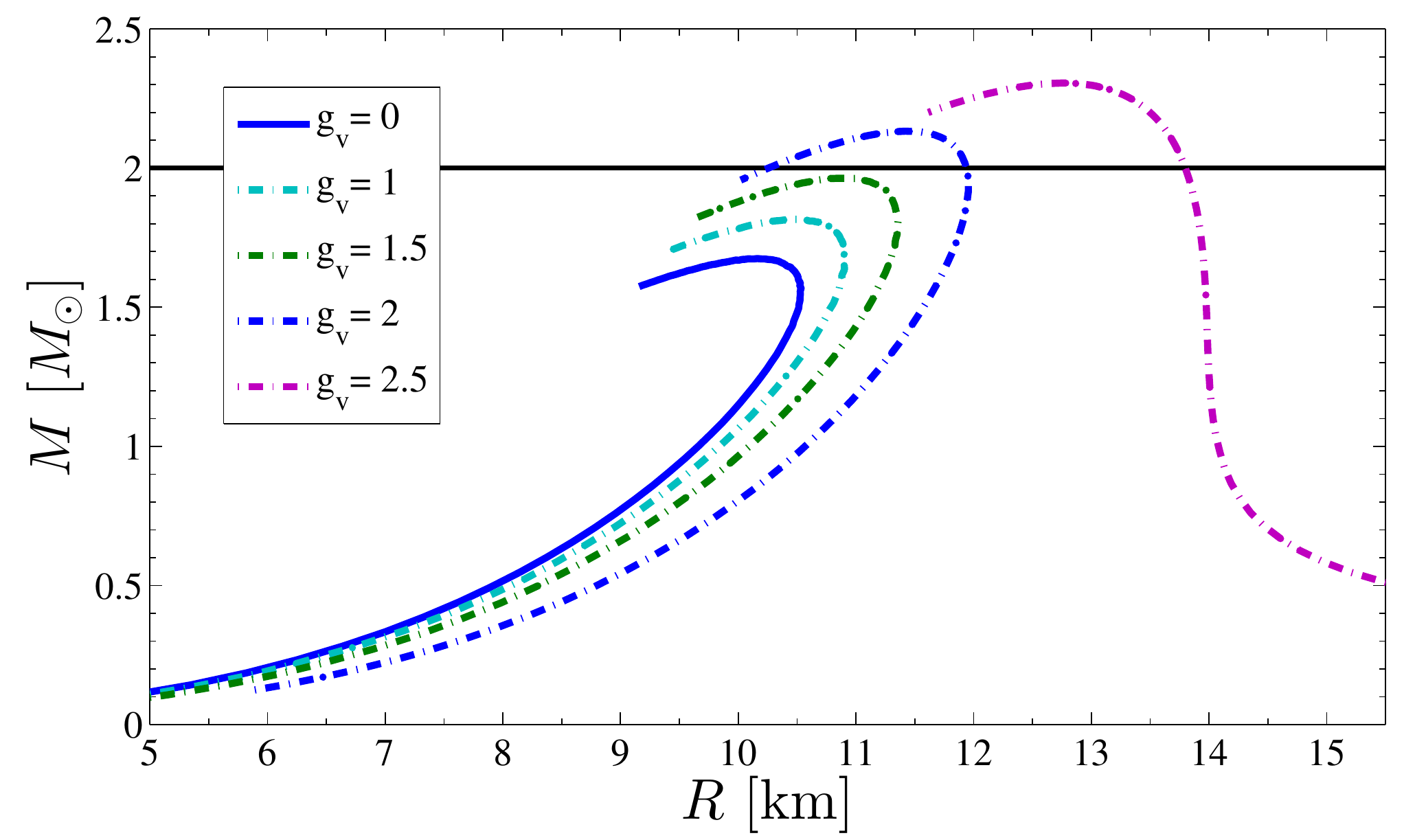}
\caption{Dependence of the mass-radius relations on the strength of the Yukawa coupling $g_v$ between quarks and the vector meson in the eLSM model.}
\label{fig:MR_gw}
\end{figure}

Finally, in Fig.~\ref{fig:MR_gw} we show the influence of the repulsive interaction on the mass-radius relation obtained in the eLSM. With increasing vector coupling the EoS becomes stiffer, and more massive and larger stable stars can be attained. For $g_v=2$ the star sequence is in the permitted radius window at $M=2$~$M_\odot$, and the largest mass is $\sim 2.15$~$M_\odot$. Beyond a certain value of the coupling the pressure becomes positive for all positive values of the energy density, which results in star sequences that contain large stars with small masses. Qualitatively similar results were reported in \cite{Zacchi:2015lwa}. 

\section{Conclusions \label{sec:concl}}

We employed the zero-temperature EoS obtained with some approximations in the eLSM to determine the mass-radius relation of compact stars, assumed to consist of matter described by this model, and compared the resulting mass and radius values to those given by the two-flavor Walecka model and the three-flavor non-interacting quark model. The mass-radius sequence obtained in the eLSM without repulsive interaction mediated by a vector meson is close to that emerging from a Walecka model which includes the self-interaction of scalar mesons, but contrary to the EoS of that model, it can not reach the desired $2$~$M_\odot$ mass value. Including the repulsive interaction in the eLSM model makes the EoS stiff enough to support in some narrow range of the Yukawa coupling compact stars with masses larger than $2$~$M_\odot$ and in the radius window of $11.0$-$12.5$~km at $M=2$~$M_\odot$, suggested by previous studies.

In the future, we would like to go beyond the mean-field approximation, used for the mesons in the eLSM, in a way that takes into account the effect of fermions in the mesonic fluctuations. At lowest order, this can be done by expanding to quadratic order the fermionic determinant obtained after integrating out the quark fields in the partition function and performing the Gaussian integral over the mesonic fields. In order to have a physically more reliable description, we also plan to include the charge neutrality and $\beta$-equilibrium conditions and improve the treatment of the interaction between vector mesons and quarks employed here.

\vspace{6pt} 

\acknowledgments{We thank P. P\'osfay for interesting discussions and the academic editor for guidance on the treatment of the neutron star crust in the Walecka model. János Takátsy was supported by the ÚNKP-18-2 New National Excellence Program of the Ministry of Human Capacities. P.~Kovács and Zs.~Szép acknowledge support by the János Bolyai Research Scholarship of the Hungarian Academy of Sciences.}


\reftitle{References}
\externalbibliography{yes}
\bibliography{Zimanyi}

\end{document}